\documentclass[floats,floatfix,showpacs,amssymb,prd,twocolumn,superscriptaddress,nofootinbib,nolongbibliography,reprint,preprintnumbers]{revtex4-1}

\usepackage{amssymb,amsmath,verbatim,mathtools,needspace,enumitem,etoolbox,graphicx,physics,microtype,afterpage,xspace,tabularx,lmodern,multirow,braket}
\usepackage{gensymb}
\usepackage[normalem]{ulem}
\usepackage[dvipsnames, usenames]{xcolor}
\definecolor{linkcolor}{rgb}{0.0,0.3,0.5}
\usepackage[unicode, colorlinks=true, linkcolor=linkcolor, citecolor=linkcolor, filecolor=linkcolor, urlcolor=linkcolor, linktocpage, breaklinks]{hyperref}
\usepackage[all]{hypcap}
\usepackage{subfigure}
\usepackage[T1]{fontenc}
\usepackage[utf8]{inputenc}
\usepackage[usenames,dvipsnames]{xcolor}
\hypersetup{colorlinks=true,citecolor=romared,linkcolor=romared,urlcolor=romared}

\definecolor{romared}{RGB}{142,0,28}

\newcommand{\be}{\begin{equation}}
\newcommand{\ee}{\end{equation}}

\def\be{\begin{equation}}
\def\ee{\end{equation}}
\newcommand{\beq}{\begin{eqnarray}}
\newcommand{\eeq}{\end{eqnarray}}

\usepackage{aas_macros}
\usepackage{makecell}
\usepackage{soul}
\usepackage{amssymb}
\usepackage{longtable}

\usepackage{lipsum}

\newcolumntype{Y}{>{\centering\arraybackslash}X}

\begin{document}

\title{Unstable chords and destructive resonant excitation of black hole quasinormal modes}

\author{Naritaka Oshita}
\affiliation{Center for Gravitational Physics and Quantum Information, Yukawa Institute for Theoretical Physics, Kyoto University, 606-8502, Kyoto, Japan}
\affiliation{The Hakubi Center for Advanced Research, Kyoto University,
Yoshida Ushinomiyacho, Sakyo-ku, Kyoto 606-8501, Japan}
\affiliation{RIKEN iTHEMS, Wako, Saitama, 351-0198, Japan}
\author{Emanuele Berti}
\affiliation{Department of Physics and Astronomy, Johns Hopkins University,
3400 N. Charles Street, Baltimore, Maryland, 21218, USA}
\author{Vitor Cardoso} 
\affiliation{Center of Gravity, Niels Bohr Institute, Blegdamsvej 17, 2100 Copenhagen, Denmark}
\affiliation{CENTRA, Departamento de F\'{\i}sica, Instituto Superior T\'ecnico -- IST, Universidade de Lisboa -- UL, Avenida Rovisco Pais 1, 1049-001 Lisboa, Portugal}

\preprint{YITP-25-44, RIKEN-iTHEMS-Report-25}

\begin{abstract}
The quasinormal mode spectrum of black holes is unstable against small modifications of the radial potential describing massless perturbations. We study how these small modifications affect the convergence of the quasinormal mode expansion and the mode excitation by computing the mode amplitudes from first principles, without relying on any fitting procedure.
We show that the decomposition of the prompt ringdown waveform is not unique: small modifications in the radial potential produce new quasinormal mode ``basis sets'' that can improve the convergence of the quasinormal mode expansion, even capturing the late-time tail.
We also study avoided crossings and exceptional points of the Kerr and Kerr-de Sitter spectrum. We show that while the mode amplitude can be resonantly excited, modes that exhibit avoided crossing destructively interfere with each other, so that the prompt ringdown waveform remains stable.
\end{abstract}

\maketitle

\noindent \textbf{\em Introduction.}
The so-called ``ringdown'' is the relaxation of a perturbed black hole (BH) to its final state via a superposition of characteristic damped exponentials with discrete frequencies: the BH's quasinormal modes (QNMs)~\cite{Kokkotas:1999bd,Berti:2009kk}. Gravitational wave detectors can measure these frequencies by observing the ringdown radiation produced in the aftermath of a BH merger. This program is known as BH spectroscopy~\cite{Dreyer:2003bv,Berti:2005ys,Baibhav:2023clw}. The QNM frequency spectrum is known to be unstable under small deformations of the effective potential of the governing dynamical equation~\cite{Nollert:1996rf,Nollert:1998ys,Barausse:2014tra,Daghigh:2020jyk,Jaramillo:2020tuu,Cheung:2021bol,Cardoso:2024mrw} or under changes in the boundary conditions~\cite{Cardoso:2016rao,Mark:2017dnq,Cardoso:2019rvt,Cardoso:2024mrw}. Despite this frequency-domain instability, the prompt time-domain ringdown waveform is stable, i.e., it is affected only perturbatively by changes in the potential or in the boundary conditions~\cite{Barausse:2014tra,Cardoso:2019rvt,Berti:2022xfj,Cardoso:2024mrw}.

Motivated by these spectral instability results, in this Letter we ask the following question: is the QNM spectrum in GR the most adequate basis set to describe the prompt time-domain ringdown?
The problem we have in mind can be illustrated with a simple analogy. Consider a string instrument (for example, a guitar). Just like a black hole, the guitar is not a conservative system, since the mechanical energy in the strings is dissipated through sound waves in the air. Now, imagine playing the guitar in an echo chamber -- a hard-wall venue such that the system is almost perfectly conservative. The QNMs of the guitar are very different from the modes of the new, conservative system. What does a listener perceive? Should we expand the sound waves in terms of the QNMs of the guitar, or in terms of the normal modes of the guitar in the venue? Which ``basis'' is more appropriate or useful?

Throughout this Letter, we use geometrical units ($G=c=1$) and we set $2M =1$, where $M$ is the BH mass.

\noindent \textbf{\em Setup.}
Using the Newman-Penrose formalism, gravitational perturbations in the background of a spinning BH of mass $M$ and dimensionless angular momentum $j = a/M = J/M^2$ can be reduced to the study of a single master wavefunction~\cite{Teukolsky:1973ha}. 
The wavefunction can be factorized into an angular part that satisfies the spin-weighted spheroidal wave equation~\cite{Berti:2005gp} and a radial part. We will consider a variant of the radial wavefunction (the Sasaki-Nakamura wavefunction $X_{\ell m \omega}$~\cite{Sasaki:1981sx,Hughes:2000pf}) which, in Fourier space, satisfies the equation
\begin{equation}
\frac{d^2 X_{\ell m \omega}}{dr_{\ast}^2} + \left(\frac{1}{2} \frac{d{\cal F}_{\ell m \omega}{}}{d r_*} - \frac{{\cal F}_{\ell m \omega}{}^2}{4} - {\cal U}_{\ell m \omega} \right) X_{\ell m \omega} = 0,
\end{equation}
where $r_*$ is the tortoise coordinate defined as $dr_*/dr = (r^2 +a^2)/(r^2 -r+a^2)$ in terms of the usual Boyer-Lindquist coordinate $r$, $\omega$ is the Laplace variable, $(\ell, m)$ label the multipolar components, and the explicit form of ${\cal F}_{\ell m \omega}$ and ${\cal U}_{\ell m \omega}$ is given in Ref.~\cite{Sasaki:1981sx}.
In the $j\to 0$ limit, the above equation reduces to the Regge-Wheeler equation.

\begin{figure}[t]
\centering
\includegraphics[width=1\linewidth]{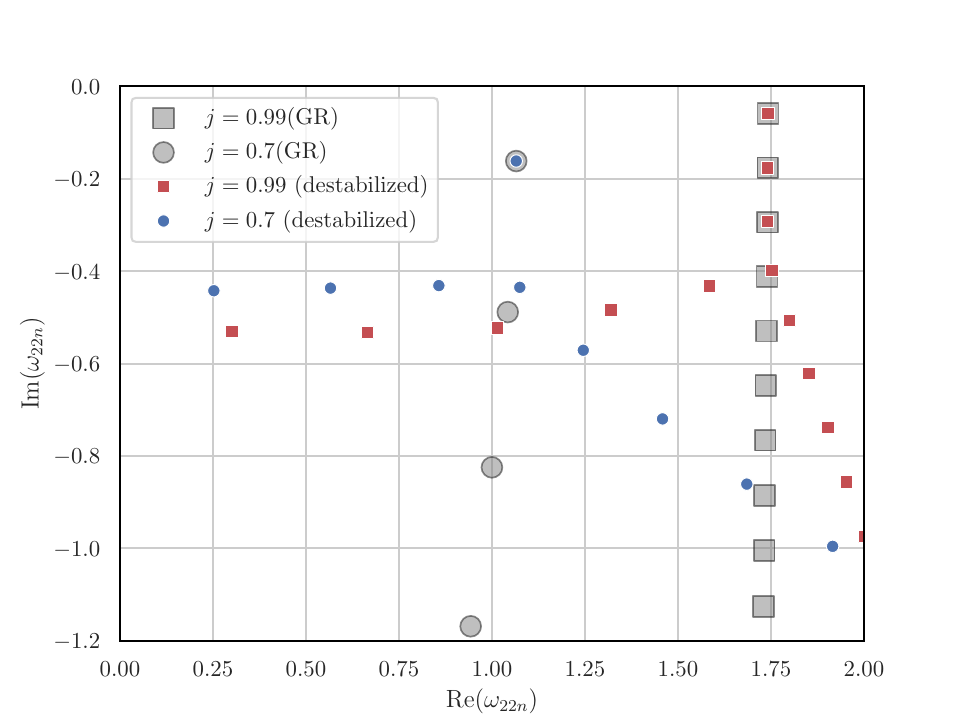}
\caption{Gray markers are the $\ell = m =2$ QNM frequencies of the Sasaki-Nakamura equation in GR for $j=0.7$ (circles) and $0.99$ (squares).
  Blue circles and red squares correspond to the QNM frequencies for a P\"oschl-Teller perturbation with $V_0 = 10^{-4}$, $x_0 =10$, and $\sigma =1$ in Eq.~\eqref{reflectivity_PT}.
}
\label{fig_destabilized_KerrQNM}
\end{figure}

The QNM frequencies are the discrete complex values of $\omega = \omega_{\ell m n}$ such that $X_{\ell m \omega}$ is purely ingoing at the horizon and purely outgoing at spatial infinity. Here the overtone number $n = 0,\,1,\,2\dots$ sorts the frequencies by the absolute value of their imaginary part.

The QNM frequency spectrum can be destabilized by small changes to the effective potential ${\cal U}_{\ell m \omega}$~\cite{Berti:2022xfj,Cardoso:2024mrw}), or even by ``soft'' changes in the boundary conditions~\cite{Cardoso:2016rao,Mark:2017dnq,Cardoso:2019rvt}. For definiteness, we choose the latter possibility and consider the following modified boundary condition:
\begin{align}
X_{\ell m \omega} =
\begin{cases}
e^{-i k_{\rm H} r_*} \ \ (r_* \to - \infty)\,,\\
e^{i \omega r_*} + \epsilon (\omega) e^{-i \omega (r_*-2x_0)}, \ \ (r_* \to \infty)\label{eq_bcs}
\end{cases}
\end{align}
where $k_{\rm H} \equiv \omega - m \Omega_{\rm H}$, $\Omega_{\rm H}$ is the horizon frequency, $\epsilon (\omega)$ is the reflectivity of the reflective ``bump,'' and $x_0$ is its radial position in terms of the tortoise coordinate.

These boundary conditions are an effective description of spacetime deformations which would induce a change in the local effective potential.
The reflective nature of the spacetime deformations is encoded in the parameter $\epsilon$ (in vacuum general relativity, $\epsilon=0$).
For example, a bump of magnitude $\delta V$ can be modeled by choosing $\epsilon \ll 1$ for $\omega^2 \gg \delta V$, and $\epsilon \simeq 1$ for $\omega^2 \ll \delta V$.
This prescription gives results identical to those obtained by explicitly incorporating a bump-like perturbation into the master equation~\cite{Cheung:2021bol}, provided that $\epsilon (\omega)$ is appropriately chosen and the bump is far enough that it does not overlap with the peak of the potential barrier.
Even when these conditions are not fully satisfied, this method remains practical for demonstrating QNM instabilities.

For definiteness, we choose the function $\epsilon (\omega)$ such that it yields the reflectivity of the P\"oschl-Teller potential $V_{\rm PT}= V_0/\cosh^2 [(r_* - x_0)/\sigma]$ in the limit $x_0 \to \infty$:
\begin{equation}
\epsilon(\omega) = \frac{\Gamma(a_+) \Gamma(a_-) \Gamma(c-a_+-a_-)}{\Gamma(a_++a_--c) \Gamma (c-a_+) \Gamma (c-a_-)}\,,\label{reflectivity_PT}
\end{equation}
with $a_{\pm} \equiv \left[ \sigma \pm \sqrt{\sigma^2 - 4 V_0} -2i \omega \right]$ and $c\equiv 1-i \omega/\sigma$.
Other potentials can be accommodated by changing $\epsilon (\omega)$.

The QNM frequencies with the new boundary conditions are shown and compared with the GR spectrum in Fig.~\ref{fig_destabilized_KerrQNM}. It is clear at a glance that the GR spectrum is unstable under small changes in the boundary conditions.

\begin{figure}[t]
\centering
\includegraphics[width=1\linewidth]{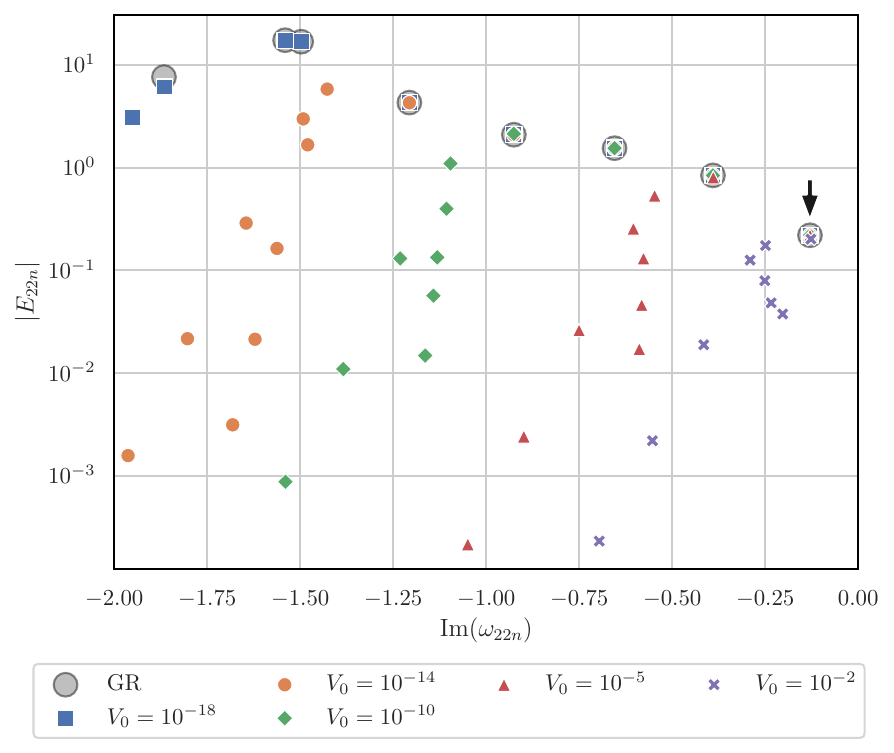}
\caption{
Absolute value of the QNMEFs $|E_{22n}|$ in GR and for destabilized modes with $j=0.9$, $x_0 = 10$, $\sigma = 1$ and different values of $V_0$, plotted as functions of ${\rm Im}(\omega_{22n})$. The QNMEF of the fundamental mode $(2,2,0)$ is marked by an arrow.}
\label{fig_FM}
\end{figure}
%

\noindent \textbf{\em Excitation and convergence of destabilized spectra.}
We have seen that small disturbances produce large changes in the spectrum. However, the prompt time domain signals is robust, as we demonstrate below (see also~\cite{Barausse:2014tra,Berti:2022xfj,Cardoso:2024mrw}). This suggests that an expansion in QNMs is not the unique (or even the most adequate) basis decomposition for the prompt signal. Consider again the string instrument analogy: at early times, when a listener receives sound waves produced by the guitar strings that did not have time to interact with the venue walls, what is the appropriate expansion? There are several possible QNM basis sets, and the convergence of the QNM expansion may be different for each set. The convergence of the expansion is relevant to model waveforms with a superposition of QNMs and to understand the ringdown starting time~\cite{Andersson:1996cm,Oshita:2024wgt}. To address this question, we study the amplitude of destabilized QNMs in a ringdown waveform $h_{{\rm G}, \ell m}(t)$ sourced by a Dirac delta function~\cite{Oshita:2024wgt}:
\begin{equation}
h_{{\rm G}, \ell m}(t) = \frac{1}{2 \pi} 
 \int d\omega \frac{{\cal A}_{{\rm out},\ell m}}{ {\cal A}_{{\rm in},\ell m}} e^{-i \omega t}\,,
\label{eq_hg}
\end{equation}
where ${\cal A}_{{\rm out},\ell m}$ and ${\cal A}_{{\rm in},\ell m}$ are the asymptotic amplitudes of the homogeneous solution to the Sasaki-Nakamura equation $R_{\ell m \omega}^{\rm in}$, which satisfies
\begin{align}
R_{\ell m \omega}^{\rm in} = 
\begin{cases}
e^{-i k_{\rm H} r_*}&r_* \to - \infty\,,\\
{\cal A}_{{\rm in},\ell m} e^{-i \omega r_*} + {\cal A}_{{\rm out},\ell m} e^{i \omega r_*}&r_* \to +\infty\,.
\end{cases}
\end{align}
(See~\cite{Rosato:2024arw} for a similar study in spherical symmetry, where however the number of QNMs was not sufficient to carefully study the convergence properties of the expansion.)

The QNM amplitude in $h_{\rm G}(t)$ can be expressed in terms of the quantities $E_{\ell mn} = -2 i \omega_{\ell mn} B_{\ell mn}$, where the $B_{\ell m n}$'s are defined as
\begin{equation}
B_{\ell m n} = \frac{{\cal A}_{\rm out}(\omega_{\ell m n})}{2 \omega_{\ell mn} {\cal A}_{\rm in}'(\omega_{\ell m n})}\,.
\end{equation}
In what follows, we will refer to the $E_{\ell m n}$'s as the ``QNM excitation factors'' (QNMEFs). We will keep the subscripts $\ell, m$ in $\omega_{\ell m n}$ and $E_{\ell m n}$, but omit them in other quantities for brevity.

We can reconstruct the waveform $h_{\rm G}$ through a superposition of QNMs with amplitude $E_{\ell m n}$ as follows:
\begin{equation}
h_{\rm E} = \sum_{n}^{N_{\rm P}} E_{\ell m n} e^{-i \omega_{\ell m n} t} - \sum_{n}^{N_{\rm R}} E_{\ell -m n}^* e^{i \omega_{\ell -m n}^* t}\,,
\label{eq_he}
\end{equation}
where the first and second term are associated to prograde and retrograde modes, respectively.
We will compute the QNMEFs using Green's function techniques~\cite{Leaver:1986gd,Sun:1988tz,Nollert:1998ys,Leaver:1985ax,Leaver:1986vnb,Andersson:1995zk,Berti:2006wq,Zhang:2013ksa,Oshita:2021iyn}, following the methods described in Ref.~\cite{Zhang:2013ksa,Oshita:2021iyn}.

\begin{figure}[t]
\centering
\includegraphics[width=1\linewidth]{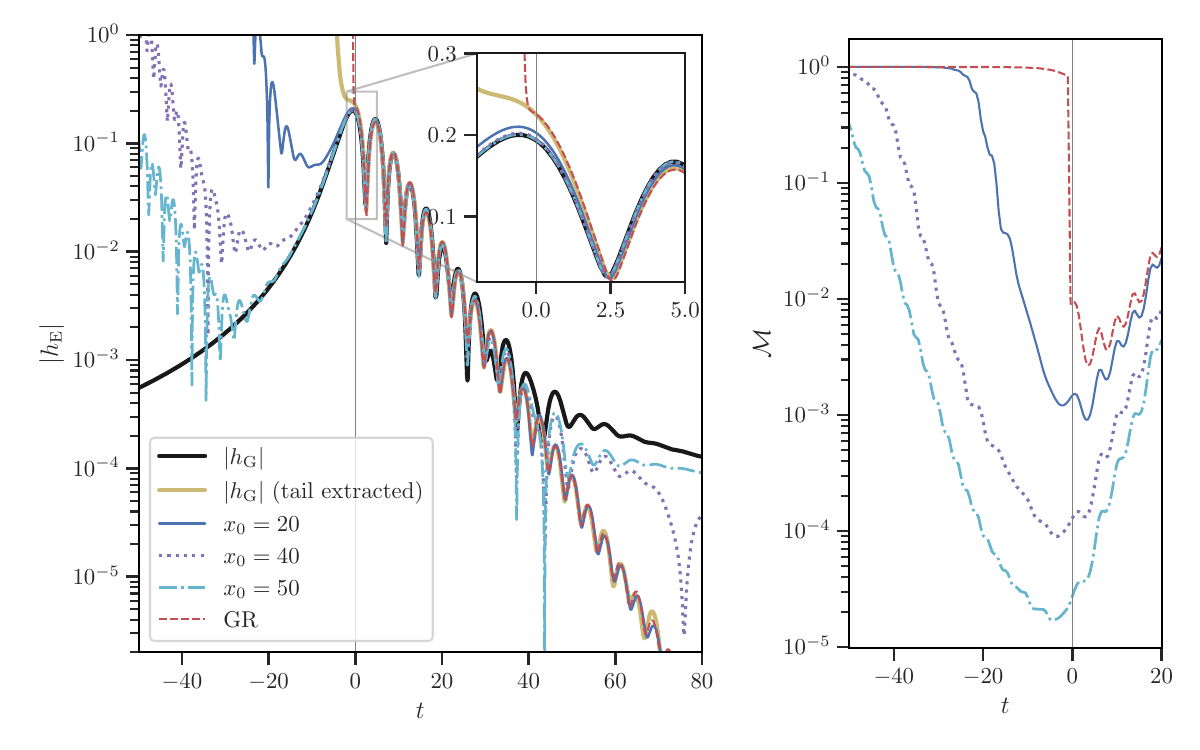}
\caption{
Left: Comparison between the ringdown $h_{\rm G}$ excited by a delta-function source in GR, and the superposition of destabilized QNM, $h_{\rm E}$, for $\ell =m=2$. 
We set $j=0.7$, $V_0 = 10^{-4}$, $\sigma = 1$, and vary $x_0$.
Right: The mismatch ${\cal M}$ between $h_{\rm G}$ and $h_{\rm E}$,
computed for $t_{\rm max} = 50$ in Eq.~(\ref{mismatch}).
}
\label{fig_recon}
\end{figure}

The QNMEFs of the destabilized spectrum are shown in Fig.~\ref{fig_FM}. 
It is apparent from the figure that the QNMEFs are also unstable under small changes of the boundary conditions. When we impose the usual vacuum boundary conditions in GR, the absolute value of the QNMEFs for $j=0.9$ has a peak at $(\ell,m,n)=(2,2,5)$ with $|E_{225}| \sim {\cal O}(10)$, while the fundamental mode has $|E_{220}| \sim {\cal O}(0.1)$. 
As we increase $V_0$ in the boundary condition~\eqref{eq_bcs}, from Fig.~\ref{fig_FM} we see that the QNMEFs for higher overtones are significantly suppressed, and the fundamental mode becomes more dominant.

Despite the spectral instability of the QNM frequencies and of the QNMEFs, the time-domain signal, as shown in Fig.~\ref{fig_recon}, is stable.
We reconstruct the signal with a superposition of QNMs whose QNMEFs satisfy $|E_{22n}| > 10^{-4}$ (this criterion is common to all cases considered in this Letter).
The reconstruction works well in vacuum GR. Less trivially, the prompt ringdown in the GR waveform $h_{\rm G}$ is very well reconstructed also when we consider QNMs with the modified boundary conditions~\eqref{eq_bcs} in $h_{\rm E}$.

Remarkably, the reconstruction using the destabilized QNMs works {\em better} than the sum of the QNMs in vacuum GR at early times ($t<0$, where the zero is defined as the earliest time at which the QNM expansion is convergent: see the right panel of Fig.~\ref{fig_recon} and Ref.~\cite{Oshita:2024wgt}).

We can also ask if we can reconstruct the power-law tail in $h_{\rm G}$ at late times with the destabilized QNMs. In the Supplemental Material we explain how we can fit full waveform (thick solid black line in Fig.~\ref{fig_recon}) to extract and subtract the contribution from the power-law tail, thus producing the solid yellow line. The inset of Fig.~\ref{fig_recon} shows a zoom-in around the peak. We see that the effect of the power-law tail is observed even around $t=0$: near the peak, the QNM expansion in GR (red dashed line) fails to capture the black solid line (which includes the contribution of the power law), but it does reconstruct the thick solid yellow (representing $h_{\rm G}$ after removal of the tail).
Perhaps the most interesting aspect of Fig.~\ref{fig_recon} is that the destabilized QNMs work remarkably well at reproducing not only the early time signal at $t<0$, but also the late-time power-law tail. For example, for $t>40$ the power-law behavior observed in the full waveform (thick solid black line) is remarkably similar to the behavior of the dash-dotted cyan line corresponding to a destabilized spectrum with $x_0=50$.

In the right panel of Fig.~\ref{fig_recon} we also show the mismatch ${\cal M} = 1- \left| \braket{h_{\rm G} | h_{\rm E}}/\sqrt{\braket{h_{\rm G} | h_{\rm G}} \braket{h_{\rm E} | h_{\rm E}}} \right|$ between $h_{\rm G}$ and $h_{\rm E}$ as a function of time, where
\begin{align}
\braket{ F | G } \equiv \int_{t}^{t_{\rm max}} dt' F(t') G^* (t')\,.
\label{mismatch}
\end{align}
The number $N_{\rm P}\,(N_{\rm R})$ of prograde (retrograde) modes in $h_{\rm E}$ is large enough to ensure convergence of ${\cal M}$.
Note that the destabilized QNMs improve the convergence of the QNM expansion at both $t<0$ and at very late times, where the signal is dominated by a power-law tail. 
This is reasonable as the destabilized QNM set consists of a growing number of trapped modes with long decay rates as $x_0$ grows, and the exponential enhancement of QNM amplitudes at earlier times is milder. Heuristically, the set of trapped modes becomes more and more similar to an infinite-dimensional normal mode basis as $x_0 \to \infty$, and this improves the convergence of the QNM expansion.

In the Supplemental Material we show that this result is supported by the simple, analytically tractable model of two rectangular potential barriers. The toy model supports our main conclusions: the expansion in terms of the destabilized QNMs exhibits better convergence at earlier times. In our analogy, an expansion on the basis set of the coupled conservative system (guitar plus room) has better convergence properties than an expansion using only the QNMs of the guitar.

Note that the boundary condition~\eqref{eq_bcs}, as well as ``bump-like'' corrections to the potential, induces very late-time echoes in the waveforms. However, as long as these changes occur in the far region ($x_0 \gg 1$), these features will only appear at late times $t \sim 2 x_0$, and they will not affect the prompt ringdown signal.

\begin{figure}[t]
\centering
\includegraphics[width=1\linewidth]{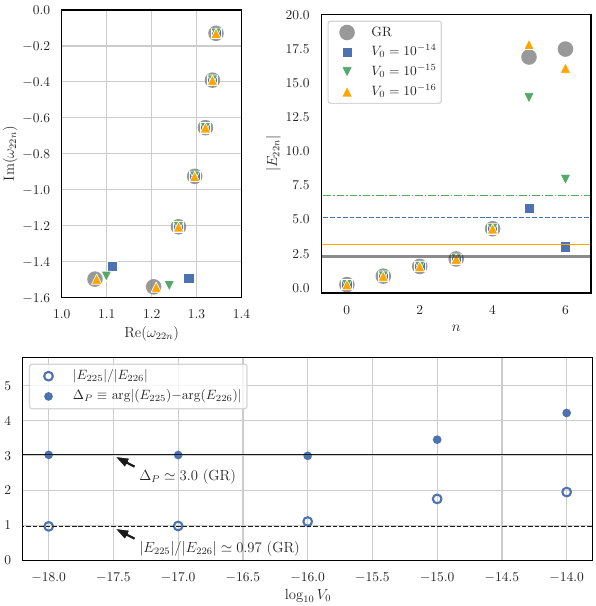}
\caption{
  Upper Left: Destabilized $(2,2,n)$  QNMs for $j=0.9$, $x_0 = 10$, $\sigma = 1$, and three values of $V_0$.
Upper Right: Absolute value of $E_{22n}$.
Horizontal lines indicate $|E_{225} + E_{226}|$ with $V_0 = 0$ (thick solid gray), $V_0 =10^{-16}$ (thin solid yellow), $V_0 =10^{-15}$ (dot-dashed green), and $V_0 =10^{-14}$ (dashed blue).
Bottom: Phase difference $\Delta_{\rm P} \equiv |\text{arg}(E_{225}) - \text{arg}(E_{226})|$ (filled blue) and ratio $|E_{225}/E_{226}|$ (hollow blue).
In GR, $\Delta_{\rm P} \simeq 3.0 \simeq \pi$ (solid black), and $|E_{225}/E_{226}| \simeq 0.97 \simeq 1$ (dashed gray). 
}
\label{fig_QNM56}
\end{figure}
%

\noindent \textbf{\em Exceptional points and QNM interference.}
Our approach can also inform us on the behavior of the resonances and exceptional points, that occur when two QNM frequencies get very close to each other or coincide exactly. The first example identified in the Kerr QNM spectrum is the $(2,2,5)$ overtone, which exhibits an anomalous dependence on the BH spin for $j \sim 0.9$~\cite{Onozawa:1996ux}. 
Interestingly, the QNMEFs of the $(2,2,5)$ and $(2,2,6)$ modes have local maxima for $j \sim 0.9$~\cite{Oshita:2021iyn} (see also Fig.~\ref{fig_FM}). These maxima were understood in terms of resonant excitation due to the avoided crossing of the two QNMs~\cite{Motohashi:2024fwt}.
Here we ask: is this resonant excitation sensitive to changes in the boundary conditions or in the effective potential?

As shown in Fig.~\ref{fig_QNM56}, the $(2,2,5)$ and $(2,2,6)$ Kerr QNMs are very sensitive to the boundary conditions. Indeed, they are destabilized already for $V_0$ in the range $[10^{-18},\,10^{-14}]$.
It is even more interesting that the $(2,2,5)$ and $(2,2,6)$ QNMEFs have nearly equal absolute values and almost opposite phases~\cite{Oshita:2021iyn}: $|E_{225}| \simeq |E_{226}|$ and $|\text{arg} (E_{225})- \text{arg} (E_{226})| \simeq \pi$ (bottom panel). 
When we introduce the small correction~\eqref{eq_bcs} to the boundary conditions, this near-perfect destructive interference between the QNMEFs disappears, but the excitation of the QNMEFs is also greatly suppressed (top right panel of Fig.~\ref{fig_QNM56}).
\begin{figure}[t]
\centering
\includegraphics[width=0.98\linewidth]{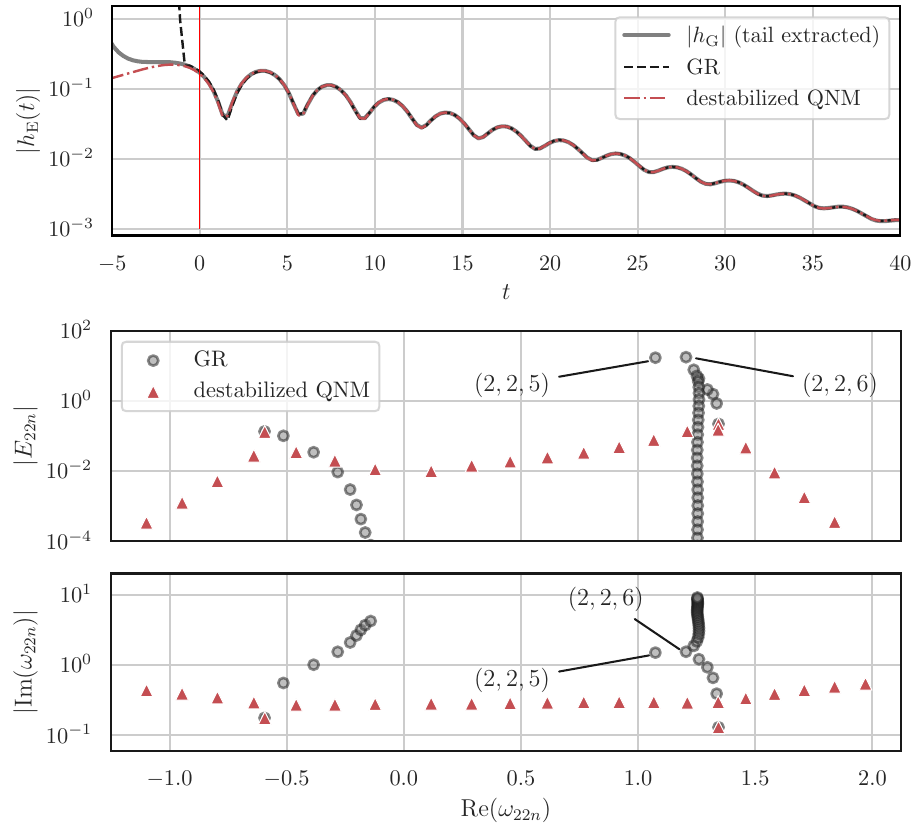}
\caption{
Top: QNM superposition $h_{\rm E}$ with Kerr QNMs (dashed black) and with destabilized QNMs (dot-dashed red).
The waveform $h_{\rm G}$ without the power-law tail (thick solid gray line) is also shown as a reference. 
We set $j = 0.9$ and $\ell = m=2$.
Center: Absolute value of the QNMEFs $|E_{22n}|$ as a function of ${\rm Re}(\omega_{22n})$.
Bottom: QNM frequencies $\omega_{22n}$ included in $h_{\rm E}$ for Kerr BHs (black open circles) and for destabilized Kerr BHs (red squares) with $V_0 = 10^{-5}$, $x_0 = 20$ and $\sigma = 1$.
}
\label{fig_stableringdown}
\end{figure}

While the local maximum of the  $(2,2,5)$ and $(2,2,6)$ QNMEFs suggests that they may be excited to a high amplitude, the destructive interference that we find here implies that they are not (or at least, not in a trivial way). Indeed, in Fig.~\ref{fig_stableringdown} we confirm that the prompt ringdown waveform constructed by superposing all QNMs does not show obvious signatures of resonant QNM excitation near the exceptional point, with $\text{max} (|E_{22n}|) \sim {\cal O} (10)$ for GR ($\epsilon=0$) and $\text{max} (|E_{22n}|) \sim {\cal O}(1)$ for destabilized spectra. Interestingly, the top panel of Fig.~\ref{fig_stableringdown} shows that the waveform found by superposing the QNMs is only mildly affected by changes in the boundary conditions.

As an additional test, we have investigated (to our knowledge, for the first time) the destructive interference of QNMs in the Kerr-de Sitter spacetime. As we show in Fig.~\ref{fig_KDS} of the Supplemental Material, the significant excitation of the $(2,2,5)$ and $(2,2,6)$ overtones corresponds to the existence of a Kerr-de Sitter exceptional point at $(j, \Lambda) \simeq (0.896, 0.034)$. However, even at the exceptional point the superposition of the two QNMEFs is rather insensitive to the resonance, and there is no apparent excitation when we consider the sum  $E_{225} + E_{226}$.

We can clarify the amplification of the QNMEFs $E_{225}$ and $E_{226}$ and their destructive interference with a simple analytical argument.
Consider two QNM frequencies $\omega_i$ ($i=1,\,2$) that are close to each other, so that $\delta \omega \equiv \omega_2 - \omega_1$ satisfies the condition $|\delta \omega| \ll |\omega_1| \simeq |\omega_2|$.
The analytic expression of the QNMEFs is $E_i = \left. A_{\rm out} (\omega)/W'(\omega) \right|_{\omega = \omega_i}$, where $A_{\rm out} (\omega)$ is the asymptotic outgoing amplitude of the in-mode solution and $W(\omega) \eqqcolon  f(\omega) (\omega- \omega_1) (\omega- \omega_2)$ is the Wronskian appearing in the denominator of the BH's Green's function.

The QNMEFs $E_i$ take the form
\begin{align}
E_1 = \frac{A_{\rm out} (\omega_1)}{f(\omega_1) (-\delta \omega)}\,,
\
E_2 = \frac{A_{\rm out} (\omega_2)}{f(\omega_2) (+\delta \omega)}\,.
\end{align}
Expanding $E_2$ for small $\delta \omega$, we find
\begin{equation}
E_2 \simeq -E_1 -E_1 \delta \omega \frac{d}{d\omega} \log (A_{\rm out}/f)|_{\omega = \omega_1}\,,
\end{equation}
and therefore $E_1 + E_2 \simeq  d \left(A_{\rm out}/f \right)/d\omega|_{\omega = \omega_1}$,
which is independent of $\delta \omega$.
This confirms that the two QNMEFs exhibiting avoided crossing interfere destructively at the onset of the QNM excitation, so that the prompt ringdown waveform in the time domain is stable.

\noindent \textbf{\em Discussion.}
In this Letter we argued that while BHs are intrinsically dissipative systems, by placing them in a cavity, we can disentangle some of the challenging aspects of their spectrum due to dissipation.
We have computed the QNM amplitudes from first principles (i.e., without fitting), considering ringdown excitation by a Dirac delta function.
We have argued that: (i) the QNM expansion of the ringdown is not unique; (ii) the reconstruction of the time-domain signal from QNMEFs is stable, even if the QNM frequencies and the QNMEFs themselves are unstable; (iii) the destabilized set of QNMs has {\em better} convergence properties in the prompt ringdown stage, because the cavity produces a large set of low-frequency, long-lived modes that look more and more like {\em normal} modes; (iv) the destabilized QNMs can be used to reconstruct the time-domain signal including very early times and late-time power-law tails; (v) the intricate destructive interference among (destabilized) QNMs near resonances or exceptional points ensures the stability of the prompt time-domain signal. In other words, systems with completely different spectra can lead to the same intermediate-time behavior. 

It is very interesting that these stability properties hold even near ``exceptional points''~\cite{Kato:101545,2019NatMa..18..783O,Wiersig:20,Ding:2022juv} -- i.e., for BH parameters such that multiple QNM frequencies are close to each other (but not quite equal, because of avoided crossings~\cite{PhysRevE.61.929,Heiss:2012dx}) and the QNMEFs are significantly enhanced~\cite{Oshita:2021iyn,Motohashi:2024fwt}.
We have shown numerically and analytically that the QNMEF enhancement at avoided crossing and at ``exceptional points'' (where two QNM frequencies are exactly equal) is affected by destructive interference, and therefore the prompt ringdown signal is only mildly affected by this phenomenon.
We have also found that the large enhancement of QNMEFs near exceptional points is significantly reduced in the presence of small perturbations, so that the prompt ringdown remains stable.

\noindent \textbf{\em Acknowledgments.}
%
We thank Yiqiu Yang and the participants of the workshop ``Ringdown Inside and Out,'' organized by the members of the Niels Bohr Institute, for helpful discussions.
The Center of Gravity is a Center of Excellence funded by the Danish National Research Foundation under grant No. 184.
N.O. sincerely thanks the members of Johns Hopkins University for their hospitality and inspiring discussions, which were invaluable for completing this work.
N.O. is supported by Japan Society for the Promotion of Science (JSPS) KAKENHI Grant No.~JP23K13111 and by the Hakubi project at Kyoto University.
E.B. is supported by NSF Grants No. AST-2307146, PHY-2207502, PHY-090003 and PHY-20043, by NASA Grant No. 21-ATP21-0010, by the John Templeton Foundation Grant 62840, by the Simons Foundation, and by the Italian Ministry of Foreign Affairs and International Cooperation grant No.~PGR01167.
V.C. acknowledges support by VILLUM Foundation (grant no.\ VIL37766) and the DNRF Chair program (grant no.\ DNRF162) by the Danish National Research Foundation.
V.C. acknowledges financial support provided under the European Union’s H2020 ERC Advanced Grant “Black holes: gravitational engines of discovery” grant agreement no.\ Gravitas–101052587. 
Views and opinions expressed are however those of the author only and do not necessarily reflect those of the European Union or the European Research Council. Neither the European Union nor the granting authority can be held responsible for them.
This project has received funding from the European Union's Horizon 2020 research and innovation programme under the Marie Sk{\l}odowska-Curie grant agreement No 101007855 and No 101131233.

\section*{Supplemental Material}

\noindent
\textbf{\em Double square barrier.}
Here we consider the QNMs of a simple double square barrier toy model:
\begin{equation}\label{sup_rec_eq}
\left( \frac{d^2}{dx^2} +\omega^2 - V_{\rm rec}(x) \right) \psi(x) = 0\,,
\end{equation}
where the potential
\begin{align}
V_{\rm rec} (x) =
\begin{cases}
0 & (x < 0)\,,\\
V_0 & (0 \leq x < d)\,,\\
0 & (d \leq x < d+b)\,,\\
\delta V & (d+b \leq x < d+b+\delta)\,,\\
0 & (d+b+\delta \leq x)\,,
\end{cases}
\end{align}
has a primary barrier of height $V_0$ and width $d$, and a (small) secondary barrier of height $\delta V \ll V_0$ and width $\delta$, located at distance $b$ from the primary barrier.
The in-mode homogeneous solution of Eq.~(\ref{sup_rec_eq}) can be found analytically:
\begin{equation}
\psi_{\rm in} = 
\begin{cases}
e^{-i\omega x}\,,  &x \to -\infty\\
B_{\rm out} (\omega) e^{i\omega x} + B_{\rm in} (\omega) e^{-i\omega x}\,,  &x \to +\infty
\end{cases}
\,.
\end{equation}
\begin{figure}[h]
\centering
\includegraphics[width=0.9\linewidth]{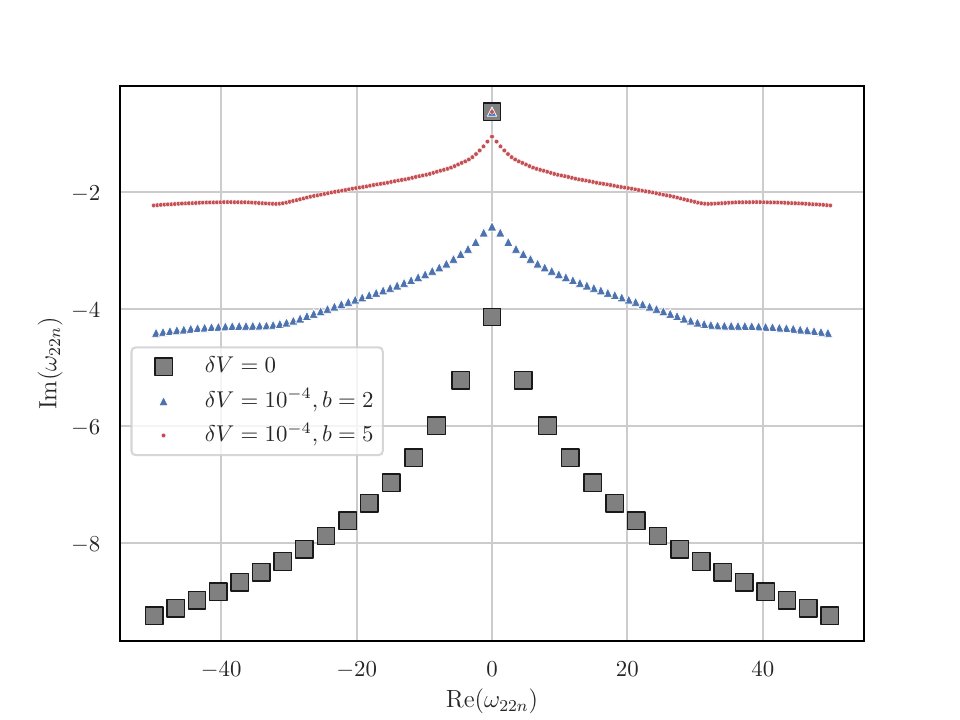}
\caption{
Complex QNM frequencies for $V_0 = 1$, $d=1$, $\delta =0.1$.
Gray squares show the QNM frequencies for a single potential barrier ($\delta V =0$). The other two QNM sets are modes destabilized by a secondary barrier with $\delta V = 10^{-4}, b=2$ (blue triangles) and $\delta V = 10^{-4}, b=5$ (little red dots).
}
\label{fig_rec_QNM}
\end{figure}
The QNM frequencies $\omega = \omega_n$ and QNMEFs, defined by 
\begin{equation}
E_n = \left.\frac{B_{\rm out}(\omega)}{2 \omega B_{\rm in}'(\omega)}\right|_{\omega = \omega_n}\,,
\end{equation}
can also be found analytically. The QNMs are significantly destabilized by tuning the parameters of the small barrier, as shown in Fig.~\ref{fig_rec_QNM}.

\begin{figure}[t]
\centering
\includegraphics[width=0.9\linewidth]{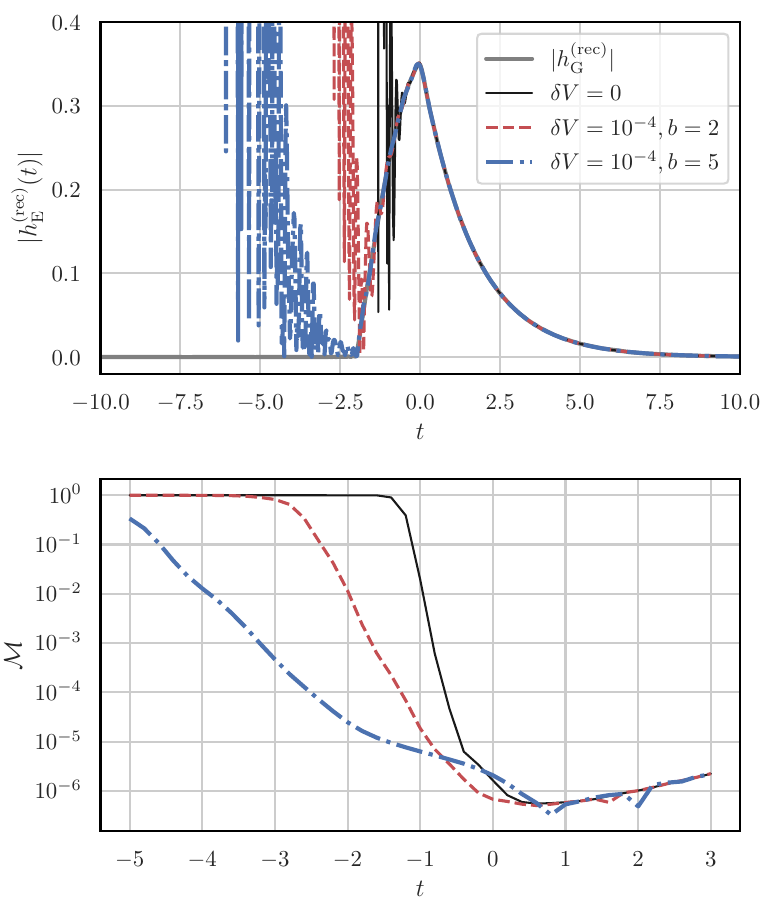}
\caption{
Reconstruction of the waveform $h_{\rm G}^{\rm (rec)}$ with the superposed QNM waveform $h_{\rm E}^{\rm (rec)}$. 
We set $V_0 = 1$, $d=1$, $\delta =0.1$, and vary $\delta V$ and $b$.
The waveform $h_{\rm E}^{\rm (rec)}$ with $\delta V = 0$ (black solid), $\delta V = 10^{-4}$, $b =2$ (red dashed) and $\delta V = 10^{-4}$, $b =5$ (blue dot-dashed) are shown.
}
\label{fig_recon_box}
\end{figure}

We can also compute a waveform $h_{\rm G}^{\rm (rec)} (t)$
\begin{equation}
h_{\rm G}^{\rm (rec)} = \frac{1}{2 \pi} \int_{-\infty}^{\infty} d \omega \frac{B_{\rm out}}{2 i \omega B_{\rm in}} e^{-i\omega t}\,,
\end{equation}
corresponding to the unperturbed single primary barrier (i.e.,  $\delta V= 0$), and compare it with $h_{\rm E}^{\rm (rec)}$:
\begin{equation}
h_{\rm E}^{\rm (rec)} = \sum_{n=0}^{N_{\rm max}'} E_{n} e^{-i \omega_n t} + 2 \text{Re} \left[\sum_{n=0}^{N_{\rm max}} E_{n} e^{-i \omega_{n} t} \right]\,,
\label{eq_rec_he}
\end{equation}
where the first term accounts for the contribution from purely imaginary QNMs, and the second term accounts for the QNMs with nonzero real parts as well as their mirror modes.
We have checked that the number of QNMs included in $h_{\rm E}^{\rm (rec)}$ is large enough to achieve convergence (as measured by the mismatch ${\cal M}$) at $t \gtrsim 2$.
We include QNMs whose QNMEFs satisfy $|E_{22n}| > 10^{-4}$ and $|\text{Re} (\omega_{22n})| < 50$ in \eqref{eq_rec_he}.
In Fig.~\ref{fig_recon_box} we show that a superposition of destabilized QNMs can reproduce the waveform $h_{\rm G}^{\rm (rec)}$ with a mismatch ${\cal M} \lesssim 10^{-5}$ at $t>0$,
and that QNM sets with long-lived trapped modes (red dashed and blue dot-dashed lines) exhibit better convergence than the QNMs of the unperturbed barrier (black solid line), with the convergence improving as $b$ increases.
This is qualitatively consistent with the Kerr QNM results discussed in the main text.
\begin{figure}[t]
\centering
\includegraphics[width=0.8\linewidth]{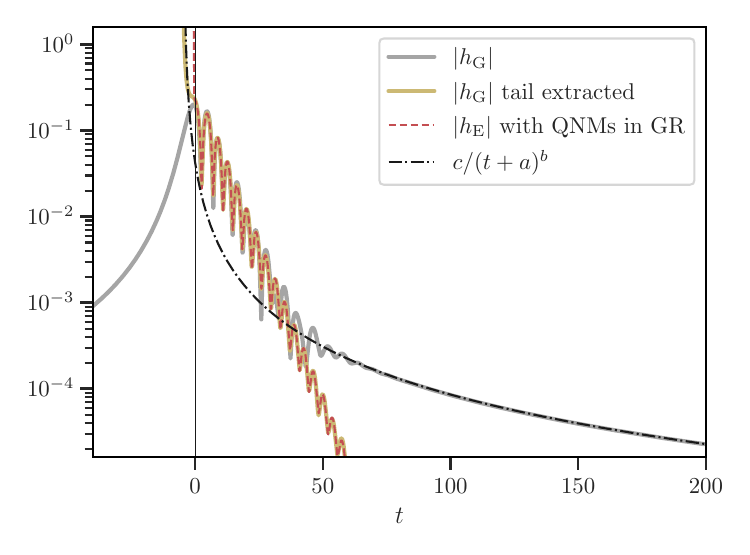}
\caption{
Fit of the model function \eqref{fit_func_tail} to extract the power-law tail from $h_{\rm G}$. The best-fit tail with $a = 4.5$, $b = 1.9716$, $c = 0.81754$ (dot-dashed black) is shown.
}
\label{fig_tailfit}
\end{figure}

\noindent
\textbf{\em Extraction of the power-law tail.}
To extract the power-law tail from $h_{\rm G}$ (thick solid yellow in the left panel in Fig.~\ref{fig_recon}), we use the following fitting function:
\begin{equation}
h_{\rm tail} (t) = \frac{c}{(t+a)^{b}}\,,
\label{fit_func_tail}
\end{equation}
where $a$, $b$ and $c$ are the fitting parameters.
We fit $h_{\rm G}$ with Eq.~\eqref{fit_func_tail} at late times ($100 \leq t \leq 200$) with \texttt{Mathematica}'s built-in function \texttt{NonlinearModelFit}.
With three fitting parameters, $\{a,b,c\}$, we find the \texttt{NonlinearModelFit} fit to be unstable.
To address this, we fix the parameter $\{a\}$, determine the remaining two parameters $\{b,c\}$ with \texttt{NonlinearModelFit}, and obtain the value of the \texttt{EstimatedVariance}.
We then systematically vary $\{a\}$ over the range of $[1,10]$ and obtain the values of $\{a,b,c\}$ that minimize the \texttt{EstimatedVariance}.
For $j = 0.7$, we find that the best-fit model for the power-law tail in $h_{\rm G}$ is given by $a \simeq 4.5$, $b \simeq 1.97$ and $c \simeq 0.818$, with \texttt{EstimatedVariance} $\sim 10^{-16}$ (see Fig.~\ref{fig_tailfit}).
We also observed that the late-time tail in the imaginary part of $h_{\rm G}$ is of the order of $10^{-6}$--$10^{-7}$, which is much smaller than the real part, so we neglect the contribution of the imaginary part of the tail.

\begin{figure*}[t]
\centering
\includegraphics[width=0.8\linewidth]{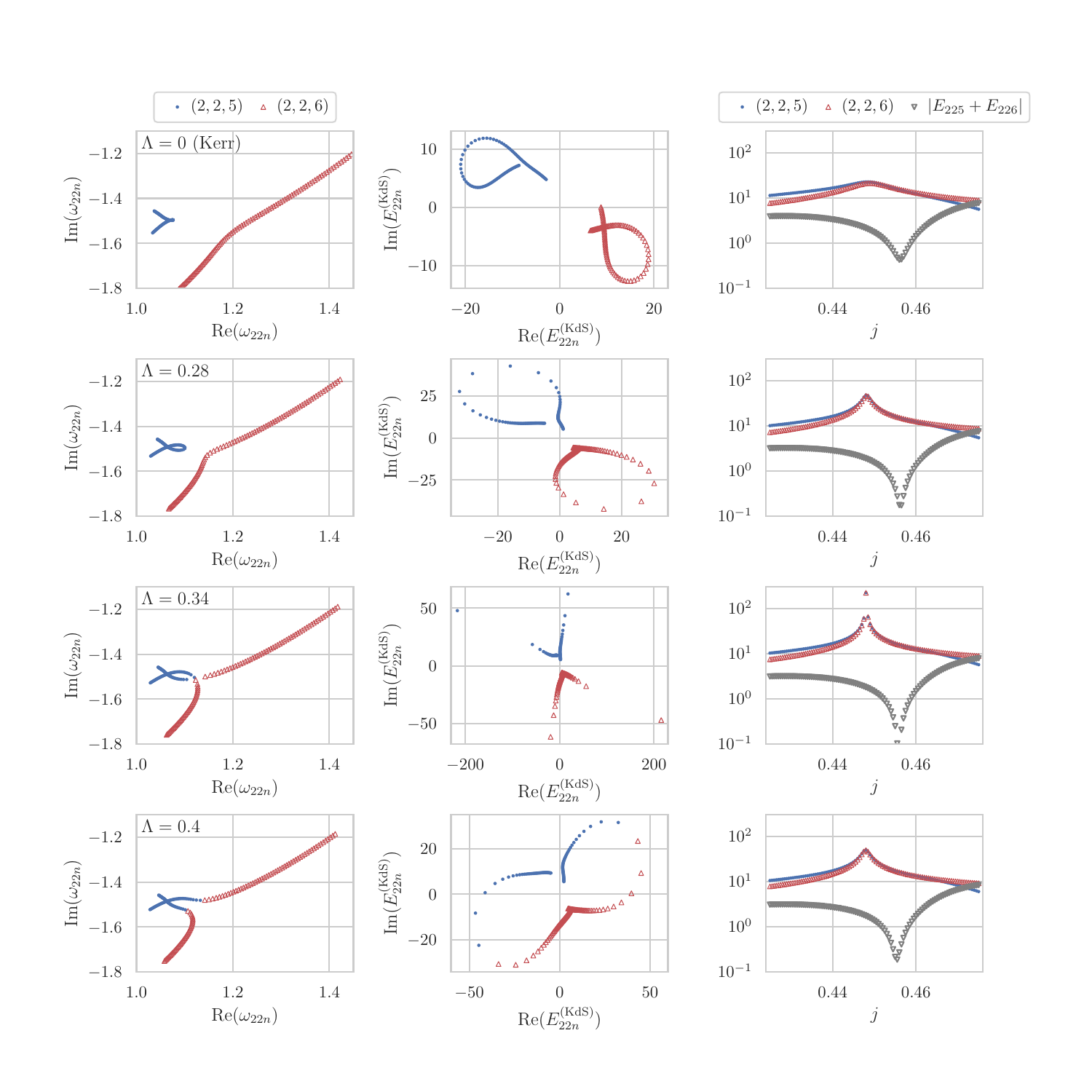}
\caption{
Avoided crossing and mode crossing (exceptional point) between the $(2,2,5)$ and $(2,2,6)$ QNMs in the Kerr-de Sitter spacetime. 
We vary the spin parameter $j$ in the range $[0.85, 0.95]$ and we consider four values of the cosmological constant: $\Lambda = 0$ (top row), $\Lambda = 0.028$ (second row), $\Lambda = 0.034$ (third row), and $\Lambda = 0.04$ (bottom row).
}
\label{fig_KDS}
\end{figure*}

\noindent
\textbf{\em Destructive resonant excitation in Kerr-de Sitter BHs.}
We compute the QNMEFs for the Kerr-de Sitter BH, defined as 
\begin{equation}
    E_{\ell m n}^{\rm (KdS)} = \frac{A^{\rm (out)}_{\rm \ell m} (\omega_{\ell mn})}{2 \omega_{\ell mn}^3 A^{\rm (in)}_{\rm \ell m} {}'(\omega_{\ell mn})}\,,
\end{equation}
from the general Heun's function that gives the analytic solution of the Teukolsky equation \cite{Suzuki:1998vy,Hatsuda:2020sbn}.
The asymptotic amplitudes we used in this calculation, $A^{\rm (in)}_{\ell m}$ and $A^{\rm (out)}_{\rm \ell m}$, are defined in Appendix~C of~\cite{Oshita:2021iyn}.

The large enhancement of the QNMEFs for $(2,2,5)$ and $(2,2,6)$ found in Ref.~\cite{Oshita:2021iyn}, and explained in terms of avoided crossing in Ref.~\cite{Motohashi:2024fwt}, corresponds to the existence of an exceptional point at $(j, \Lambda) \simeq (0.896, 0.034)$. As shown in Fig.~\ref{fig_QNM56}, we find that there is significant destructive interference even at the {\it onset} of QNM excitation for all values of $\Lambda$, including $\Lambda = 0$ (i.e., the Kerr case).


\begin{thebibliography}{41}%
\makeatletter
\providecommand \@ifxundefined [1]{%
 \@ifx{#1\undefined}
}%
\providecommand \@ifnum [1]{%
 \ifnum #1\expandafter \@firstoftwo
 \else \expandafter \@secondoftwo
 \fi
}%
\providecommand \@ifx [1]{%
 \ifx #1\expandafter \@firstoftwo
 \else \expandafter \@secondoftwo
 \fi
}%
\providecommand \natexlab [1]{#1}%
\providecommand \enquote  [1]{``#1''}%
\providecommand \bibnamefont  [1]{#1}%
\providecommand \bibfnamefont [1]{#1}%
\providecommand \citenamefont [1]{#1}%
\providecommand \href@noop [0]{\@secondoftwo}%
\providecommand \href [0]{\begingroup \@sanitize@url \@href}%
\providecommand \@href[1]{\@@startlink{#1}\@@href}%
\providecommand \@@href[1]{\endgroup#1\@@endlink}%
\providecommand \@sanitize@url [0]{\catcode `\\12\catcode `\$12\catcode
  `\&12\catcode `\#12\catcode `\^12\catcode `\_12\catcode `\%12\relax}%
\providecommand \@@startlink[1]{}%
\providecommand \@@endlink[0]{}%
\providecommand \url  [0]{\begingroup\@sanitize@url \@url }%
\providecommand \@url [1]{\endgroup\@href {#1}{\urlprefix }}%
\providecommand \urlprefix  [0]{URL }%
\providecommand \Eprint [0]{\href }%
\providecommand \doibase [0]{http://dx.doi.org/}%
\providecommand \selectlanguage [0]{\@gobble}%
\providecommand \bibinfo  [0]{\@secondoftwo}%
\providecommand \bibfield  [0]{\@secondoftwo}%
\providecommand \translation [1]{[#1]}%
\providecommand \BibitemOpen [0]{}%
\providecommand \bibitemStop [0]{}%
\providecommand \bibitemNoStop [0]{.\EOS\space}%
\providecommand \EOS [0]{\spacefactor3000\relax}%
\providecommand \BibitemShut  [1]{\csname bibitem#1\endcsname}%
\let\auto@bib@innerbib\@empty
\bibitem [{\citenamefont {Kokkotas}\ and\ \citenamefont
  {Schmidt}(1999)}]{Kokkotas:1999bd}%
  \BibitemOpen
  \bibfield  {author} {\bibinfo {author} {\bibfnamefont {K.~D.}\ \bibnamefont
  {Kokkotas}}\ and\ \bibinfo {author} {\bibfnamefont {B.~G.}\ \bibnamefont
  {Schmidt}},\ }\href {\doibase 10.12942/lrr-1999-2} {\bibfield  {journal}
  {\bibinfo  {journal} {Living Rev. Rel.}\ }\textbf {\bibinfo {volume} {2}},\
  \bibinfo {pages} {2} (\bibinfo {year} {1999})},\ \Eprint
  {http://arxiv.org/abs/gr-qc/9909058} {arXiv:gr-qc/9909058} \BibitemShut
  {NoStop}%
\bibitem [{\citenamefont {Berti}\ \emph {et~al.}(2009)\citenamefont {Berti},
  \citenamefont {Cardoso},\ and\ \citenamefont {Starinets}}]{Berti:2009kk}%
  \BibitemOpen
  \bibfield  {author} {\bibinfo {author} {\bibfnamefont {E.}~\bibnamefont
  {Berti}}, \bibinfo {author} {\bibfnamefont {V.}~\bibnamefont {Cardoso}}, \
  and\ \bibinfo {author} {\bibfnamefont {A.~O.}\ \bibnamefont {Starinets}},\
  }\href {\doibase 10.1088/0264-9381/26/16/163001} {\bibfield  {journal}
  {\bibinfo  {journal} {Class. Quant. Grav.}\ }\textbf {\bibinfo {volume}
  {26}},\ \bibinfo {pages} {163001} (\bibinfo {year} {2009})},\ \Eprint
  {http://arxiv.org/abs/0905.2975} {arXiv:0905.2975 [gr-qc]} \BibitemShut
  {NoStop}%
\bibitem [{\citenamefont {Dreyer}\ \emph {et~al.}(2004)\citenamefont {Dreyer},
  \citenamefont {Kelly}, \citenamefont {Krishnan}, \citenamefont {Finn},
  \citenamefont {Garrison},\ and\ \citenamefont
  {Lopez-Aleman}}]{Dreyer:2003bv}%
  \BibitemOpen
  \bibfield  {author} {\bibinfo {author} {\bibfnamefont {O.}~\bibnamefont
  {Dreyer}}, \bibinfo {author} {\bibfnamefont {B.~J.}\ \bibnamefont {Kelly}},
  \bibinfo {author} {\bibfnamefont {B.}~\bibnamefont {Krishnan}}, \bibinfo
  {author} {\bibfnamefont {L.~S.}\ \bibnamefont {Finn}}, \bibinfo {author}
  {\bibfnamefont {D.}~\bibnamefont {Garrison}}, \ and\ \bibinfo {author}
  {\bibfnamefont {R.}~\bibnamefont {Lopez-Aleman}},\ }\href {\doibase
  10.1088/0264-9381/21/4/003} {\bibfield  {journal} {\bibinfo  {journal}
  {Class. Quant. Grav.}\ }\textbf {\bibinfo {volume} {21}},\ \bibinfo {pages}
  {787} (\bibinfo {year} {2004})},\ \Eprint
  {http://arxiv.org/abs/gr-qc/0309007} {arXiv:gr-qc/0309007} \BibitemShut
  {NoStop}%
\bibitem [{\citenamefont {Berti}\ \emph
  {et~al.}(2006{\natexlab{a}})\citenamefont {Berti}, \citenamefont {Cardoso},\
  and\ \citenamefont {Will}}]{Berti:2005ys}%
  \BibitemOpen
  \bibfield  {author} {\bibinfo {author} {\bibfnamefont {E.}~\bibnamefont
  {Berti}}, \bibinfo {author} {\bibfnamefont {V.}~\bibnamefont {Cardoso}}, \
  and\ \bibinfo {author} {\bibfnamefont {C.~M.}\ \bibnamefont {Will}},\ }\href
  {\doibase 10.1103/PhysRevD.73.064030} {\bibfield  {journal} {\bibinfo
  {journal} {Phys. Rev. D}\ }\textbf {\bibinfo {volume} {73}},\ \bibinfo
  {pages} {064030} (\bibinfo {year} {2006}{\natexlab{a}})},\ \Eprint
  {http://arxiv.org/abs/gr-qc/0512160} {arXiv:gr-qc/0512160} \BibitemShut
  {NoStop}%
\bibitem [{\citenamefont {Baibhav}\ \emph {et~al.}(2023)\citenamefont
  {Baibhav}, \citenamefont {Cheung}, \citenamefont {Berti}, \citenamefont
  {Cardoso}, \citenamefont {Carullo}, \citenamefont {Cotesta}, \citenamefont
  {Del~Pozzo},\ and\ \citenamefont {Duque}}]{Baibhav:2023clw}%
  \BibitemOpen
  \bibfield  {author} {\bibinfo {author} {\bibfnamefont {V.}~\bibnamefont
  {Baibhav}}, \bibinfo {author} {\bibfnamefont {M.~H.-Y.}\ \bibnamefont
  {Cheung}}, \bibinfo {author} {\bibfnamefont {E.}~\bibnamefont {Berti}},
  \bibinfo {author} {\bibfnamefont {V.}~\bibnamefont {Cardoso}}, \bibinfo
  {author} {\bibfnamefont {G.}~\bibnamefont {Carullo}}, \bibinfo {author}
  {\bibfnamefont {R.}~\bibnamefont {Cotesta}}, \bibinfo {author} {\bibfnamefont
  {W.}~\bibnamefont {Del~Pozzo}}, \ and\ \bibinfo {author} {\bibfnamefont
  {F.}~\bibnamefont {Duque}},\ }\href {\doibase 10.1103/PhysRevD.108.104020}
  {\bibfield  {journal} {\bibinfo  {journal} {Phys. Rev. D}\ }\textbf {\bibinfo
  {volume} {108}},\ \bibinfo {pages} {104020} (\bibinfo {year} {2023})},\
  \Eprint {http://arxiv.org/abs/2302.03050} {arXiv:2302.03050 [gr-qc]}
  \BibitemShut {NoStop}%
\bibitem [{\citenamefont {Nollert}(1996)}]{Nollert:1996rf}%
  \BibitemOpen
  \bibfield  {author} {\bibinfo {author} {\bibfnamefont {H.-P.}\ \bibnamefont
  {Nollert}},\ }\href {\doibase 10.1103/PhysRevD.53.4397} {\bibfield  {journal}
  {\bibinfo  {journal} {Phys. Rev. D}\ }\textbf {\bibinfo {volume} {53}},\
  \bibinfo {pages} {4397} (\bibinfo {year} {1996})},\ \Eprint
  {http://arxiv.org/abs/gr-qc/9602032} {arXiv:gr-qc/9602032} \BibitemShut
  {NoStop}%
\bibitem [{\citenamefont {Nollert}\ and\ \citenamefont
  {Price}(1999)}]{Nollert:1998ys}%
  \BibitemOpen
  \bibfield  {author} {\bibinfo {author} {\bibfnamefont {H.-P.}\ \bibnamefont
  {Nollert}}\ and\ \bibinfo {author} {\bibfnamefont {R.~H.}\ \bibnamefont
  {Price}},\ }\href {\doibase 10.1063/1.532698} {\bibfield  {journal} {\bibinfo
   {journal} {J. Math. Phys.}\ }\textbf {\bibinfo {volume} {40}},\ \bibinfo
  {pages} {980} (\bibinfo {year} {1999})},\ \Eprint
  {http://arxiv.org/abs/gr-qc/9810074} {arXiv:gr-qc/9810074} \BibitemShut
  {NoStop}%
\bibitem [{\citenamefont {Barausse}\ \emph {et~al.}(2014)\citenamefont
  {Barausse}, \citenamefont {Cardoso},\ and\ \citenamefont
  {Pani}}]{Barausse:2014tra}%
  \BibitemOpen
  \bibfield  {author} {\bibinfo {author} {\bibfnamefont {E.}~\bibnamefont
  {Barausse}}, \bibinfo {author} {\bibfnamefont {V.}~\bibnamefont {Cardoso}}, \
  and\ \bibinfo {author} {\bibfnamefont {P.}~\bibnamefont {Pani}},\ }\href
  {\doibase 10.1103/PhysRevD.89.104059} {\bibfield  {journal} {\bibinfo
  {journal} {Phys. Rev. D}\ }\textbf {\bibinfo {volume} {89}},\ \bibinfo
  {pages} {104059} (\bibinfo {year} {2014})},\ \Eprint
  {http://arxiv.org/abs/1404.7149} {arXiv:1404.7149 [gr-qc]} \BibitemShut
  {NoStop}%
\bibitem [{\citenamefont {Daghigh}\ \emph {et~al.}(2020)\citenamefont
  {Daghigh}, \citenamefont {Green},\ and\ \citenamefont
  {Morey}}]{Daghigh:2020jyk}%
  \BibitemOpen
  \bibfield  {author} {\bibinfo {author} {\bibfnamefont {R.~G.}\ \bibnamefont
  {Daghigh}}, \bibinfo {author} {\bibfnamefont {M.~D.}\ \bibnamefont {Green}},
  \ and\ \bibinfo {author} {\bibfnamefont {J.~C.}\ \bibnamefont {Morey}},\
  }\href {\doibase 10.1103/PhysRevD.101.104009} {\bibfield  {journal} {\bibinfo
   {journal} {Phys. Rev. D}\ }\textbf {\bibinfo {volume} {101}},\ \bibinfo
  {pages} {104009} (\bibinfo {year} {2020})},\ \Eprint
  {http://arxiv.org/abs/2002.07251} {arXiv:2002.07251 [gr-qc]} \BibitemShut
  {NoStop}%
\bibitem [{\citenamefont {Jaramillo}\ \emph {et~al.}(2021)\citenamefont
  {Jaramillo}, \citenamefont {Panosso~Macedo},\ and\ \citenamefont
  {Al~Sheikh}}]{Jaramillo:2020tuu}%
  \BibitemOpen
  \bibfield  {author} {\bibinfo {author} {\bibfnamefont {J.~L.}\ \bibnamefont
  {Jaramillo}}, \bibinfo {author} {\bibfnamefont {R.}~\bibnamefont
  {Panosso~Macedo}}, \ and\ \bibinfo {author} {\bibfnamefont {L.}~\bibnamefont
  {Al~Sheikh}},\ }\href {\doibase 10.1103/PhysRevX.11.031003} {\bibfield
  {journal} {\bibinfo  {journal} {Phys. Rev. X}\ }\textbf {\bibinfo {volume}
  {11}},\ \bibinfo {pages} {031003} (\bibinfo {year} {2021})},\ \Eprint
  {http://arxiv.org/abs/2004.06434} {arXiv:2004.06434 [gr-qc]} \BibitemShut
  {NoStop}%
\bibitem [{\citenamefont {Cheung}\ \emph {et~al.}(2022)\citenamefont {Cheung},
  \citenamefont {Destounis}, \citenamefont {Macedo}, \citenamefont {Berti},\
  and\ \citenamefont {Cardoso}}]{Cheung:2021bol}%
  \BibitemOpen
  \bibfield  {author} {\bibinfo {author} {\bibfnamefont {M.~H.-Y.}\
  \bibnamefont {Cheung}}, \bibinfo {author} {\bibfnamefont {K.}~\bibnamefont
  {Destounis}}, \bibinfo {author} {\bibfnamefont {R.~P.}\ \bibnamefont
  {Macedo}}, \bibinfo {author} {\bibfnamefont {E.}~\bibnamefont {Berti}}, \
  and\ \bibinfo {author} {\bibfnamefont {V.}~\bibnamefont {Cardoso}},\ }\href
  {\doibase 10.1103/PhysRevLett.128.111103} {\bibfield  {journal} {\bibinfo
  {journal} {Phys. Rev. Lett.}\ }\textbf {\bibinfo {volume} {128}},\ \bibinfo
  {pages} {111103} (\bibinfo {year} {2022})},\ \Eprint
  {http://arxiv.org/abs/2111.05415} {arXiv:2111.05415 [gr-qc]} \BibitemShut
  {NoStop}%
\bibitem [{\citenamefont {Cardoso}\ \emph {et~al.}(2024)\citenamefont
  {Cardoso}, \citenamefont {Kastha},\ and\ \citenamefont
  {Panosso~Macedo}}]{Cardoso:2024mrw}%
  \BibitemOpen
  \bibfield  {author} {\bibinfo {author} {\bibfnamefont {V.}~\bibnamefont
  {Cardoso}}, \bibinfo {author} {\bibfnamefont {S.}~\bibnamefont {Kastha}}, \
  and\ \bibinfo {author} {\bibfnamefont {R.}~\bibnamefont {Panosso~Macedo}},\
  }\href {\doibase 10.1103/PhysRevD.110.024016} {\bibfield  {journal} {\bibinfo
   {journal} {Phys. Rev. D}\ }\textbf {\bibinfo {volume} {110}},\ \bibinfo
  {pages} {024016} (\bibinfo {year} {2024})},\ \Eprint
  {http://arxiv.org/abs/2404.01374} {arXiv:2404.01374 [gr-qc]} \BibitemShut
  {NoStop}%
\bibitem [{\citenamefont {Cardoso}\ \emph {et~al.}(2016)\citenamefont
  {Cardoso}, \citenamefont {Franzin},\ and\ \citenamefont
  {Pani}}]{Cardoso:2016rao}%
  \BibitemOpen
  \bibfield  {author} {\bibinfo {author} {\bibfnamefont {V.}~\bibnamefont
  {Cardoso}}, \bibinfo {author} {\bibfnamefont {E.}~\bibnamefont {Franzin}}, \
  and\ \bibinfo {author} {\bibfnamefont {P.}~\bibnamefont {Pani}},\ }\href
  {\doibase 10.1103/PhysRevLett.116.171101} {\bibfield  {journal} {\bibinfo
  {journal} {Phys. Rev. Lett.}\ }\textbf {\bibinfo {volume} {116}},\ \bibinfo
  {pages} {171101} (\bibinfo {year} {2016})},\ \bibinfo {note} {[Erratum:
  Phys.Rev.Lett. 117, 089902 (2016)]},\ \Eprint
  {http://arxiv.org/abs/1602.07309} {arXiv:1602.07309 [gr-qc]} \BibitemShut
  {NoStop}%
\bibitem [{\citenamefont {Mark}\ \emph {et~al.}(2017)\citenamefont {Mark},
  \citenamefont {Zimmerman}, \citenamefont {Du},\ and\ \citenamefont
  {Chen}}]{Mark:2017dnq}%
  \BibitemOpen
  \bibfield  {author} {\bibinfo {author} {\bibfnamefont {Z.}~\bibnamefont
  {Mark}}, \bibinfo {author} {\bibfnamefont {A.}~\bibnamefont {Zimmerman}},
  \bibinfo {author} {\bibfnamefont {S.~M.}\ \bibnamefont {Du}}, \ and\ \bibinfo
  {author} {\bibfnamefont {Y.}~\bibnamefont {Chen}},\ }\href {\doibase
  10.1103/PhysRevD.96.084002} {\bibfield  {journal} {\bibinfo  {journal} {Phys.
  Rev. D}\ }\textbf {\bibinfo {volume} {96}},\ \bibinfo {pages} {084002}
  (\bibinfo {year} {2017})},\ \Eprint {http://arxiv.org/abs/1706.06155}
  {arXiv:1706.06155 [gr-qc]} \BibitemShut {NoStop}%
\bibitem [{\citenamefont {Cardoso}\ and\ \citenamefont
  {Pani}(2019)}]{Cardoso:2019rvt}%
  \BibitemOpen
  \bibfield  {author} {\bibinfo {author} {\bibfnamefont {V.}~\bibnamefont
  {Cardoso}}\ and\ \bibinfo {author} {\bibfnamefont {P.}~\bibnamefont {Pani}},\
  }\href {\doibase 10.1007/s41114-019-0020-4} {\bibfield  {journal} {\bibinfo
  {journal} {Living Rev. Rel.}\ }\textbf {\bibinfo {volume} {22}},\ \bibinfo
  {pages} {4} (\bibinfo {year} {2019})},\ \Eprint
  {http://arxiv.org/abs/1904.05363} {arXiv:1904.05363 [gr-qc]} \BibitemShut
  {NoStop}%
\bibitem [{\citenamefont {Berti}\ \emph {et~al.}(2022)\citenamefont {Berti},
  \citenamefont {Cardoso}, \citenamefont {Cheung}, \citenamefont {Di~Filippo},
  \citenamefont {Duque}, \citenamefont {Martens},\ and\ \citenamefont
  {Mukohyama}}]{Berti:2022xfj}%
  \BibitemOpen
  \bibfield  {author} {\bibinfo {author} {\bibfnamefont {E.}~\bibnamefont
  {Berti}}, \bibinfo {author} {\bibfnamefont {V.}~\bibnamefont {Cardoso}},
  \bibinfo {author} {\bibfnamefont {M.~H.-Y.}\ \bibnamefont {Cheung}}, \bibinfo
  {author} {\bibfnamefont {F.}~\bibnamefont {Di~Filippo}}, \bibinfo {author}
  {\bibfnamefont {F.}~\bibnamefont {Duque}}, \bibinfo {author} {\bibfnamefont
  {P.}~\bibnamefont {Martens}}, \ and\ \bibinfo {author} {\bibfnamefont
  {S.}~\bibnamefont {Mukohyama}},\ }\href {\doibase
  10.1103/PhysRevD.106.084011} {\bibfield  {journal} {\bibinfo  {journal}
  {Phys. Rev. D}\ }\textbf {\bibinfo {volume} {106}},\ \bibinfo {pages}
  {084011} (\bibinfo {year} {2022})},\ \Eprint
  {http://arxiv.org/abs/2205.08547} {arXiv:2205.08547 [gr-qc]} \BibitemShut
  {NoStop}%
\bibitem [{\citenamefont {Teukolsky}(1973)}]{Teukolsky:1973ha}%
  \BibitemOpen
  \bibfield  {author} {\bibinfo {author} {\bibfnamefont {S.~A.}\ \bibnamefont
  {Teukolsky}},\ }\href {\doibase 10.1086/152444} {\bibfield  {journal}
  {\bibinfo  {journal} {Astrophys. J.}\ }\textbf {\bibinfo {volume} {185}},\
  \bibinfo {pages} {635} (\bibinfo {year} {1973})}\BibitemShut {NoStop}%
\bibitem [{\citenamefont {Berti}\ \emph
  {et~al.}(2006{\natexlab{b}})\citenamefont {Berti}, \citenamefont {Cardoso},\
  and\ \citenamefont {Casals}}]{Berti:2005gp}%
  \BibitemOpen
  \bibfield  {author} {\bibinfo {author} {\bibfnamefont {E.}~\bibnamefont
  {Berti}}, \bibinfo {author} {\bibfnamefont {V.}~\bibnamefont {Cardoso}}, \
  and\ \bibinfo {author} {\bibfnamefont {M.}~\bibnamefont {Casals}},\ }\href
  {\doibase 10.1103/PhysRevD.73.109902} {\bibfield  {journal} {\bibinfo
  {journal} {Phys. Rev. D}\ }\textbf {\bibinfo {volume} {73}},\ \bibinfo
  {pages} {024013} (\bibinfo {year} {2006}{\natexlab{b}})},\ \bibinfo {note}
  {[Erratum: Phys.Rev.D 73, 109902 (2006)]},\ \Eprint
  {http://arxiv.org/abs/gr-qc/0511111} {arXiv:gr-qc/0511111} \BibitemShut
  {NoStop}%
\bibitem [{\citenamefont {Sasaki}\ and\ \citenamefont
  {Nakamura}(1982)}]{Sasaki:1981sx}%
  \BibitemOpen
  \bibfield  {author} {\bibinfo {author} {\bibfnamefont {M.}~\bibnamefont
  {Sasaki}}\ and\ \bibinfo {author} {\bibfnamefont {T.}~\bibnamefont
  {Nakamura}},\ }\href {\doibase 10.1143/PTP.67.1788} {\bibfield  {journal}
  {\bibinfo  {journal} {Prog. Theor. Phys.}\ }\textbf {\bibinfo {volume}
  {67}},\ \bibinfo {pages} {1788} (\bibinfo {year} {1982})}\BibitemShut
  {NoStop}%
\bibitem [{\citenamefont {Hughes}(2000)}]{Hughes:2000pf}%
  \BibitemOpen
  \bibfield  {author} {\bibinfo {author} {\bibfnamefont {S.~A.}\ \bibnamefont
  {Hughes}},\ }\href {\doibase 10.1103/PhysRevD.62.044029} {\bibfield
  {journal} {\bibinfo  {journal} {Phys. Rev. D}\ }\textbf {\bibinfo {volume}
  {62}},\ \bibinfo {pages} {044029} (\bibinfo {year} {2000})},\ \bibinfo {note}
  {[Erratum: Phys.Rev.D 67, 089902 (2003)]},\ \Eprint
  {http://arxiv.org/abs/gr-qc/0002043} {arXiv:gr-qc/0002043} \BibitemShut
  {NoStop}%
\bibitem [{\citenamefont {Andersson}(1997)}]{Andersson:1996cm}%
  \BibitemOpen
  \bibfield  {author} {\bibinfo {author} {\bibfnamefont {N.}~\bibnamefont
  {Andersson}},\ }\href {\doibase 10.1103/PhysRevD.55.468} {\bibfield
  {journal} {\bibinfo  {journal} {Phys. Rev. D}\ }\textbf {\bibinfo {volume}
  {55}},\ \bibinfo {pages} {468} (\bibinfo {year} {1997})},\ \Eprint
  {http://arxiv.org/abs/gr-qc/9607064} {arXiv:gr-qc/9607064} \BibitemShut
  {NoStop}%
\bibitem [{\citenamefont {Oshita}\ and\ \citenamefont
  {Cardoso}(2024)}]{Oshita:2024wgt}%
  \BibitemOpen
  \bibfield  {author} {\bibinfo {author} {\bibfnamefont {N.}~\bibnamefont
  {Oshita}}\ and\ \bibinfo {author} {\bibfnamefont {V.}~\bibnamefont
  {Cardoso}},\ }\href@noop {} {\  (\bibinfo {year} {2024})},\ \Eprint
  {http://arxiv.org/abs/2407.02563} {arXiv:2407.02563 [gr-qc]} \BibitemShut
  {NoStop}%
\bibitem [{\citenamefont {Rosato}\ \emph {et~al.}(2024)\citenamefont {Rosato},
  \citenamefont {Destounis},\ and\ \citenamefont {Pani}}]{Rosato:2024arw}%
  \BibitemOpen
  \bibfield  {author} {\bibinfo {author} {\bibfnamefont {R.~F.}\ \bibnamefont
  {Rosato}}, \bibinfo {author} {\bibfnamefont {K.}~\bibnamefont {Destounis}}, \
  and\ \bibinfo {author} {\bibfnamefont {P.}~\bibnamefont {Pani}},\ }\href
  {\doibase 10.1103/PhysRevD.110.L121501} {\bibfield  {journal} {\bibinfo
  {journal} {Phys. Rev. D}\ }\textbf {\bibinfo {volume} {110}},\ \bibinfo
  {pages} {L121501} (\bibinfo {year} {2024})},\ \Eprint
  {http://arxiv.org/abs/2406.01692} {arXiv:2406.01692 [gr-qc]} \BibitemShut
  {NoStop}%
\bibitem [{\citenamefont {Leaver}(1986{\natexlab{a}})}]{Leaver:1986gd}%
  \BibitemOpen
  \bibfield  {author} {\bibinfo {author} {\bibfnamefont {E.~W.}\ \bibnamefont
  {Leaver}},\ }\href {\doibase 10.1103/PhysRevD.34.384} {\bibfield  {journal}
  {\bibinfo  {journal} {Phys. Rev. D}\ }\textbf {\bibinfo {volume} {34}},\
  \bibinfo {pages} {384} (\bibinfo {year} {1986}{\natexlab{a}})}\BibitemShut
  {NoStop}%
\bibitem [{\citenamefont {Sun}\ and\ \citenamefont {Price}(1988)}]{Sun:1988tz}%
  \BibitemOpen
  \bibfield  {author} {\bibinfo {author} {\bibfnamefont {Y.}~\bibnamefont
  {Sun}}\ and\ \bibinfo {author} {\bibfnamefont {R.~H.}\ \bibnamefont
  {Price}},\ }\href {\doibase 10.1103/PhysRevD.38.1040} {\bibfield  {journal}
  {\bibinfo  {journal} {Phys. Rev. D}\ }\textbf {\bibinfo {volume} {38}},\
  \bibinfo {pages} {1040} (\bibinfo {year} {1988})}\BibitemShut {NoStop}%
\bibitem [{\citenamefont {Leaver}(1985)}]{Leaver:1985ax}%
  \BibitemOpen
  \bibfield  {author} {\bibinfo {author} {\bibfnamefont {E.~W.}\ \bibnamefont
  {Leaver}},\ }\href {\doibase 10.1098/rspa.1985.0119} {\bibfield  {journal}
  {\bibinfo  {journal} {Proc. Roy. Soc. Lond. A}\ }\textbf {\bibinfo {volume}
  {402}},\ \bibinfo {pages} {285} (\bibinfo {year} {1985})}\BibitemShut
  {NoStop}%
\bibitem [{\citenamefont {Leaver}(1986{\natexlab{b}})}]{Leaver:1986vnb}%
  \BibitemOpen
  \bibfield  {author} {\bibinfo {author} {\bibfnamefont {E.~W.}\ \bibnamefont
  {Leaver}},\ }\href {\doibase 10.1063/1.527130} {\bibfield  {journal}
  {\bibinfo  {journal} {J. Math. Phys.}\ }\textbf {\bibinfo {volume} {27}},\
  \bibinfo {pages} {1238} (\bibinfo {year} {1986}{\natexlab{b}})}\BibitemShut
  {NoStop}%
\bibitem [{\citenamefont {Andersson}(1995)}]{Andersson:1995zk}%
  \BibitemOpen
  \bibfield  {author} {\bibinfo {author} {\bibfnamefont {N.}~\bibnamefont
  {Andersson}},\ }\href {\doibase 10.1103/PhysRevD.51.353} {\bibfield
  {journal} {\bibinfo  {journal} {Phys. Rev. D}\ }\textbf {\bibinfo {volume}
  {51}},\ \bibinfo {pages} {353} (\bibinfo {year} {1995})}\BibitemShut
  {NoStop}%
\bibitem [{\citenamefont {Berti}\ and\ \citenamefont
  {Cardoso}(2006)}]{Berti:2006wq}%
  \BibitemOpen
  \bibfield  {author} {\bibinfo {author} {\bibfnamefont {E.}~\bibnamefont
  {Berti}}\ and\ \bibinfo {author} {\bibfnamefont {V.}~\bibnamefont
  {Cardoso}},\ }\href {\doibase 10.1103/PhysRevD.74.104020} {\bibfield
  {journal} {\bibinfo  {journal} {Phys. Rev. D}\ }\textbf {\bibinfo {volume}
  {74}},\ \bibinfo {pages} {104020} (\bibinfo {year} {2006})},\ \Eprint
  {http://arxiv.org/abs/gr-qc/0605118} {arXiv:gr-qc/0605118} \BibitemShut
  {NoStop}%
\bibitem [{\citenamefont {Zhang}\ \emph {et~al.}(2013)\citenamefont {Zhang},
  \citenamefont {Berti},\ and\ \citenamefont {Cardoso}}]{Zhang:2013ksa}%
  \BibitemOpen
  \bibfield  {author} {\bibinfo {author} {\bibfnamefont {Z.}~\bibnamefont
  {Zhang}}, \bibinfo {author} {\bibfnamefont {E.}~\bibnamefont {Berti}}, \ and\
  \bibinfo {author} {\bibfnamefont {V.}~\bibnamefont {Cardoso}},\ }\href
  {\doibase 10.1103/PhysRevD.88.044018} {\bibfield  {journal} {\bibinfo
  {journal} {Phys. Rev. D}\ }\textbf {\bibinfo {volume} {88}},\ \bibinfo
  {pages} {044018} (\bibinfo {year} {2013})},\ \Eprint
  {http://arxiv.org/abs/1305.4306} {arXiv:1305.4306 [gr-qc]} \BibitemShut
  {NoStop}%
\bibitem [{\citenamefont {Oshita}(2021)}]{Oshita:2021iyn}%
  \BibitemOpen
  \bibfield  {author} {\bibinfo {author} {\bibfnamefont {N.}~\bibnamefont
  {Oshita}},\ }\href {\doibase 10.1103/PhysRevD.104.124032} {\bibfield
  {journal} {\bibinfo  {journal} {Phys. Rev. D}\ }\textbf {\bibinfo {volume}
  {104}},\ \bibinfo {pages} {124032} (\bibinfo {year} {2021})},\ \Eprint
  {http://arxiv.org/abs/2109.09757} {arXiv:2109.09757 [gr-qc]} \BibitemShut
  {NoStop}%
\bibitem [{\citenamefont {Onozawa}(1997)}]{Onozawa:1996ux}%
  \BibitemOpen
  \bibfield  {author} {\bibinfo {author} {\bibfnamefont {H.}~\bibnamefont
  {Onozawa}},\ }\href {\doibase 10.1103/PhysRevD.55.3593} {\bibfield  {journal}
  {\bibinfo  {journal} {Phys. Rev. D}\ }\textbf {\bibinfo {volume} {55}},\
  \bibinfo {pages} {3593} (\bibinfo {year} {1997})},\ \Eprint
  {http://arxiv.org/abs/gr-qc/9610048} {arXiv:gr-qc/9610048} \BibitemShut
  {NoStop}%
\bibitem [{\citenamefont {Motohashi}(2024)}]{Motohashi:2024fwt}%
  \BibitemOpen
  \bibfield  {author} {\bibinfo {author} {\bibfnamefont {H.}~\bibnamefont
  {Motohashi}},\ }\href@noop {} {\  (\bibinfo {year} {2024})},\ \Eprint
  {http://arxiv.org/abs/2407.15191} {arXiv:2407.15191 [gr-qc]} \BibitemShut
  {NoStop}%
\bibitem [{\citenamefont {Kato}(1976)}]{Kato:101545}%
  \BibitemOpen
  \bibfield  {author} {\bibinfo {author} {\bibfnamefont {T.}~\bibnamefont
  {Kato}},\ }\href {https://cds.cern.ch/record/101545} {\emph {\bibinfo {title}
  {{Perturbation theory for linear operators; 2nd ed.}}}},\ Grundlehren der
  mathematischen Wissenschaften : a series of comprehensive studies in
  mathematics\ (\bibinfo  {publisher} {Springer},\ \bibinfo {address}
  {Berlin},\ \bibinfo {year} {1976})\BibitemShut {NoStop}%
\bibitem [{\citenamefont {{{\"O}zdemir}}\ \emph {et~al.}(2019)\citenamefont
  {{{\"O}zdemir}}, \citenamefont {{Rotter}}, \citenamefont {{Nori}},\ and\
  \citenamefont {{Yang}}}]{2019NatMa..18..783O}%
  \BibitemOpen
  \bibfield  {author} {\bibinfo {author} {\bibfnamefont {{\c{S}}.~K.}\
  \bibnamefont {{{\"O}zdemir}}}, \bibinfo {author} {\bibfnamefont
  {S.}~\bibnamefont {{Rotter}}}, \bibinfo {author} {\bibfnamefont
  {F.}~\bibnamefont {{Nori}}}, \ and\ \bibinfo {author} {\bibfnamefont
  {L.}~\bibnamefont {{Yang}}},\ }\href {\doibase 10.1038/s41563-019-0304-9}
  {\bibfield  {journal} {\bibinfo  {journal} {Nature Materials}\ }\textbf
  {\bibinfo {volume} {18}},\ \bibinfo {pages} {783} (\bibinfo {year}
  {2019})}\BibitemShut {NoStop}%
\bibitem [{\citenamefont {Wiersig}(2020)}]{Wiersig:20}%
  \BibitemOpen
  \bibfield  {author} {\bibinfo {author} {\bibfnamefont {J.}~\bibnamefont
  {Wiersig}},\ }\href {\doibase 10.1364/PRJ.396115} {\bibfield  {journal}
  {\bibinfo  {journal} {Photon. Res.}\ }\textbf {\bibinfo {volume} {8}},\
  \bibinfo {pages} {1457} (\bibinfo {year} {2020})}\BibitemShut {NoStop}%
\bibitem [{\citenamefont {Ding}\ \emph {et~al.}(2022)\citenamefont {Ding},
  \citenamefont {Fang},\ and\ \citenamefont {Ma}}]{Ding:2022juv}%
  \BibitemOpen
  \bibfield  {author} {\bibinfo {author} {\bibfnamefont {K.}~\bibnamefont
  {Ding}}, \bibinfo {author} {\bibfnamefont {C.}~\bibnamefont {Fang}}, \ and\
  \bibinfo {author} {\bibfnamefont {G.}~\bibnamefont {Ma}},\ }\href {\doibase
  10.1038/s42254-022-00516-5} {\bibfield  {journal} {\bibinfo  {journal}
  {Nature Rev. Phys.}\ }\textbf {\bibinfo {volume} {4}},\ \bibinfo {pages}
  {745} (\bibinfo {year} {2022})},\ \Eprint {http://arxiv.org/abs/2204.11601}
  {arXiv:2204.11601 [quant-ph]} \BibitemShut {NoStop}%
\bibitem [{\citenamefont {Heiss}(2000)}]{PhysRevE.61.929}%
  \BibitemOpen
  \bibfield  {author} {\bibinfo {author} {\bibfnamefont {W.~D.}\ \bibnamefont
  {Heiss}},\ }\href {\doibase 10.1103/PhysRevE.61.929} {\bibfield  {journal}
  {\bibinfo  {journal} {Phys. Rev. E}\ }\textbf {\bibinfo {volume} {61}},\
  \bibinfo {pages} {929} (\bibinfo {year} {2000})}\BibitemShut {NoStop}%
\bibitem [{\citenamefont {Heiss}(2012)}]{Heiss:2012dx}%
  \BibitemOpen
  \bibfield  {author} {\bibinfo {author} {\bibfnamefont {W.~D.}\ \bibnamefont
  {Heiss}},\ }\href {\doibase 10.1088/1751-8113/45/44/444016} {\bibfield
  {journal} {\bibinfo  {journal} {J. Phys. A}\ }\textbf {\bibinfo {volume}
  {45}},\ \bibinfo {pages} {444016} (\bibinfo {year} {2012})},\ \Eprint
  {http://arxiv.org/abs/1210.7536} {arXiv:1210.7536 [quant-ph]} \BibitemShut
  {NoStop}%
\bibitem [{\citenamefont {Suzuki}\ \emph {et~al.}(1998)\citenamefont {Suzuki},
  \citenamefont {Takasugi},\ and\ \citenamefont {Umetsu}}]{Suzuki:1998vy}%
  \BibitemOpen
  \bibfield  {author} {\bibinfo {author} {\bibfnamefont {H.}~\bibnamefont
  {Suzuki}}, \bibinfo {author} {\bibfnamefont {E.}~\bibnamefont {Takasugi}}, \
  and\ \bibinfo {author} {\bibfnamefont {H.}~\bibnamefont {Umetsu}},\ }\href
  {\doibase 10.1143/PTP.100.491} {\bibfield  {journal} {\bibinfo  {journal}
  {Prog. Theor. Phys.}\ }\textbf {\bibinfo {volume} {100}},\ \bibinfo {pages}
  {491} (\bibinfo {year} {1998})},\ \Eprint
  {http://arxiv.org/abs/gr-qc/9805064} {arXiv:gr-qc/9805064} \BibitemShut
  {NoStop}%
\bibitem [{\citenamefont {Hatsuda}(2020)}]{Hatsuda:2020sbn}%
  \BibitemOpen
  \bibfield  {author} {\bibinfo {author} {\bibfnamefont {Y.}~\bibnamefont
  {Hatsuda}},\ }\href {\doibase 10.1088/1361-6382/abc82e} {\bibfield  {journal}
  {\bibinfo  {journal} {Class. Quant. Grav.}\ }\textbf {\bibinfo {volume}
  {38}},\ \bibinfo {pages} {025015} (\bibinfo {year} {2020})},\ \Eprint
  {http://arxiv.org/abs/2006.08957} {arXiv:2006.08957 [gr-qc]} \BibitemShut
  {NoStop}%
\end{thebibliography}
\end{document}